\documentstyle[12pt,epsfig]{article}

\begin{document}
\newcommand{\p}{\partial}
\newcommand{\ls}{\left(}
\newcommand{\rs}{\right)}
\newcommand{\beq}{\begin{equation}}
\newcommand{\eeq}{\end{equation}}
\newcommand{\beqa}{\begin{eqnarray}}
\newcommand{\eeqa}{\end{eqnarray}}
\newcommand{\bdm}{\begin{displaymath}}
\newcommand{\edm}{\end{displaymath}}
\newcommand{\fps}{f_{\pi}^2 }
\newcommand{\mks}{m_{{\mathrm K}}^2 }
\newcommand{\ms}{m_{{\mathrm K}}^{*} }
\newcommand{\msq}{m_{{\mathrm K}}^{*2} }
\newcommand{\rhos}{\rho_{\mathrm s} }
\newcommand{\rhob}{\rho_{\mathrm B} }
\begin{center}

~ {\large \bf   K$^+$'s Collective Flow in Heavy-ion Collisions
predicted by  Covariant Kaon Dynamics}

\vspace{1.0cm}

 XING Yong-Zhong(ÐÏÓÀÖÒ) \footnote{E-mail:
 yzxing@tsnc.edu.cn}$^{,*}$, DAI Yue-Qian$^{1}$(´÷ÔÂÜç) and ZHENG Yu-Ming(Ö£ÓñÃ÷)$^{1,2}$

\vspace{0.5cm}

1. Institute of fundamental Physics, Tianshui Normal University, Tianshui 741000, China\\
2. China Institute of Atomic Energy, P.O. Box 275(18), Beijing
102413, China\\

\end{center}
\begin{center}
{\bf Abstract}
\end{center}
\begin{small}
\begin{quote}

The directed and elliptic flows of positively charged Kaon
produced in  $^{58}_{28}$Ni + $^{58}_{28}$Ni reaction at incident
kinetic energy 1.91 AGeV, experimental data are released newly by
V. Zinyuk,et.al. in Ref.[arXiv: 1403.1504v2 [nucl-ex] 8 Apr 2014],
are reproduced  by using  the covariant kaon dynamics. Our
numerical results indicate qualitatively the Lorentz force  is
necessary to explained reasonably the data as soon as the
space-like part of kaon's vector potential is involved. The
sensitivity of $K^+$ directed as well as differential directed
flow on the Lorentz force  are also observed near target rapidity.

\end{quote}
\end{small}
{\bf PACS} number(s): 25.75.Ld, 24.10.Jv, 25.75.Dw \\
{\bf Keyword}:Kaon meson; collective flow; heavy-ion collisions;
Covariant dynamics.

\newpage

Kaon mesons production in heavy ion reactions at intermediate
energies has been one of high interest topics in nuclear physics for
several decades  since it opens the possibility to attack several
fundamental questions of nuclear and hadron physics which are not
only interesting by themselves but have also astrophysical
implications[1-6]. K$^+$ meson is produced in the early stage of
heavy ion collisions at which the nuclear density in the reaction
zone is much higher than the saturation density ($ \rho_0 =
0.16fm^{-3} $). After its production $K^{+}$ meson escapes
near-freely from the reaction zone due to the relatively low
$K^{+}N$ scattering cross section ($ \sim$ 10 milibarn ) and the
absence of the absorption channel of a $K^{+}$ meson on a nucleon in
strong interaction. So, kaon production is proposed to be a
sensitive probe to study the nuclear equation of state (EOS) in
dense hadronic matter[7]. In 2002, P. Danielewicz et.al.[8]
presented an EoS constrained by heavy ion flow data including Kaons
has found wide-spread attention, meanwhile they also stated that it
is impossible to find a unique formulation of the EoS that
reproduces all the data. Other published experimental works on flow
comparing data with various transport codes come to similar
conclusions[9,10]: not only does one fail to reproduce all the data
with a given code, there is unfortunately so far no really
satisfactory agreement between the various theoretical approaches.
In view of such situation, recently, W. Reisdorf and his
Collaboration[11]  started a large effort to complement earlier
works by a systematic investigation encompassing a large range of
system-energies using the large acceptance apparatus FOPI at the SIS
accelerator. Most recently, they[12] reported on the simultaneous
measurements of kaon and antikaon mesons in Ni + Ni collisions at an
incident beam energy of 1.91AGeV that is close to the various
strangeness production threshold energies. These newly-delivered
experimental data provide us a chance to enrich our knowledge on the
complicated dynamical processes and to improve the existed
theoretical model. In practice, in Ref.[12] the authors have already
compared the data to the predictions
 of some transport models for the azimuthal anisotropy
of K$^-$ mesons in a wide range of rapidity and the centrality
dependence of $p_{T}$ differential flow of K$^+$ mesons.

As have already known[13,14,15], the combination of Quantum
Molecular Dynamics(QMD) with the covariant kaon dynamics, in which a
Lorentz-like force can be derived for the kaon mesons inside the
nuclear medium, is one of successful theoretical transport model for
simulating the K$^+$ mesons production in heavy ion collisions at
the SIS energy region. With this model the collective flow of $K^+$
meson and some associated produced particles have been reproduced
reasonably[15,16,17]. Naturally, a question arises that is whether
the model could  describe the collective flow of $K^+$'s produced in
this reaction. Alteratively, can the covariant kaon dynamics model
reproduce the newly released experimental data? Obviously, to answer
this question make sense since the validity of the model can been
checked by virtue of the new data. Therefore, we present here the
theoretical calculation within the covariant kaon dynamics and the
comparison of them with the data. To this end we first review
briefly the model and then give the results and discussions.

In the QMD model[13,18,19] each nucleon is represented by a coherent
state of the form (we set $\hbar$,c=1)

\begin{equation}
\psi(\vec{r},\vec{p}_0,t)= \frac{exp[i\vec{p}_0\cdot
(\vec{r}-\vec{r}_0)]}{(2{\pi}L)^{3/4}}e^{-(\vec{r}-\vec{r}_0)^{2}/4L}.
\label{1}
\end{equation}
where $\vec{r}_0$ is the time dependent center of the Gaussian wave
packet in coordinate space. The width L is kept constant, which
means that one does not allow the spreading of the wave function.
Otherwise, the whole nucleus would spread in coordinate space as a
function of time. L is set to be L=1.08 $fm^{2}$ corresponding to a
root mean square radius of the nulceons of 1.8 fm. To keep the
formulation as close as possible to the classical transport theory,
one uses Wigner density instead of working with wave function. The
wigner representation of our Gaussian wave packets obeys the
uncertainty relation $\Delta r_{x}\Delta p_{x}=\hbar/2$.

The time evolution of the N-body distribution is determined by the
motion of the centroids of the Gaussians (${ r_{i0}},{ p_{i0}}$),
which are propagated by the Poisson brackets
\begin{equation}
  \dot{\vec{p}}_{i0} = \{\vec{r}_{i0},H\} {\,}~~~\dot{\vec{r}}_{i0} = \{\vec{p}_{i0},H\}.
\label{2}
\end{equation}
Where $H$ is the nuclear Hamiltonian
\begin{equation}
  H = \sum_{i}\sqrt{{{\bf p}_{i0}}^2+m_i^2}+\frac{1}{2}\sum_{i\neq
  j}(U_{ij}^{str}+U_{ij}^{cou}).
\label{3}
\end{equation}
here $U_{ij}^{str}$ is the nuclear mean field,
$U_{ij}^{cou}$ is the Coulomb interaction.

The natural framework to study the interaction between pseudoscalar
mesons and baryons at low energies is the chiral perturbation theory
(ChPT). From the chiral Lagrangian the field equations for the
$K^\pm$--mesons are derived from the Euler-Lagrange equations
[13,19]

 \beq \left[ \partial_\mu
\partial^\mu \pm \frac{3i}{4f_{\pi}^{*2}} j_\mu \partial^\mu +
\left( \mks - \frac{\Sigma_{\mathrm{KN}}}{f_{\pi}^{*2}} \rhos
\right) \right] \phi_{\mathrm{K^\pm}} (x) = 0 \quad . \label{4} \eeq

Here the mean field approximation has already been applied. In Eq.
(4) $ \rhos $ is the baryon scalar density, $j_\mu$ is the baryon
four-vector current, $f_{\pi}^{*}$ is the in-medium pion decay
constant. Introducing the kaonic vector potential \beq V_\mu =
\frac{3}{8f_{\pi }^{*2}} j_\mu , \label{5} \eeq Eq. (4) can be
rewritten in the form[21] \beq \left[ \left(
\partial_\mu \pm i V_\mu \right)^2  + \msq \right]
\phi_{\mathrm{K^\pm}} (x) = 0~. \label{6} \eeq Thus, the vector
field is introduced by minimal coupling into the Klein-Gordon
equation. The effective mass $\ms$ of the kaon is then given
by[21,22]  \beq \ms = \sqrt{ \mks -
\frac{\Sigma_{\mathrm{KN}}}{f_{\pi }^{*2}} \rhos
     + V_\mu V^\mu }
\quad . \label{7} \eeq where $m_K $ = 0.496 GeV is the bare kaon
mass. Due to the bosonic character, the coupling of the scalar field
to the mass term is no longer linear as for the baryons but
quadratic and contains an additional contribution originating from
the vector field. The effective quasi-particle mass defined by Eq.
(7) is a Lorentz scalar and is equal for $K^+$ and $K^-$.

The $K^{\pm}$ single-particle energy is expressed as \beq
 \omega_{K^{\pm}}({\bf k},\rho)  =  \sqrt{{\bf k}^{*2} +  m_{K}^{*2}} \pm V_0
\label{8} \eeq where $k^{*} = k \mp V $ is the kaon effective
momentum, $k_{\mu} = (k_{0}, {\bf k})$, $V_{\mu} = (V_{0}, {\bf
V})$. The kaon vector field is introduced by minimal coupling into
the Klein-Gordon with opposite signs for $K^{+}$ and $K^{-}$.
$m_{K}^{*}$ is the kaon effective (Dirac) mass. The kaon (antikaon)
potential $U_{K^{\pm}}({\bf k},\rho)$ is defined as \beq
 U_{K^{\pm}}({\bf k},\rho)  =  \omega_{K^{\pm}}({\bf k},\rho) -
 \omega_0({\bf k}),
\label{9} \eeq where \beq \omega_0({\bf k})  =   \sqrt{{\bf k}^2 +
\mks }. \label{10} \eeq

In nuclear matter at rest the spatial components of the vector
potential vanish, i.e. ${\bf V} = 0$, and Eqs. (6) reduce to the
expression already given in Ref.[20]. The kaon (antikaon) potential
 $U_{K^{\pm}}({\bf k}, \rho)$ then reduce to the form
\beq
 U_{K^{\pm}}({\bf k}, {\bf V} = 0, \rho)
  =  \sqrt{{\bf k}^2 +  \mks - \frac{\Sigma_{\mathrm{KN}}}{f_{\pi}^{*2}} \rhos
     + V_{0}^2 } \pm V_0 - \sqrt{{\bf k}^2 +  \mks }.
\label{11} \eeq

The covariant equations of motion are obtained in the classical
(testparticle) limit from the relativistic transport equation for
the kaons which can be derived from Eqs. (6). They are analogous to
the corresponding relativistic equations for baryons and read[13,21]
\beqa \frac{ d  q^\mu}{d\tau} = \frac{k^{*\mu}}{\ms} \quad , \quad
\frac{ d  k^{*\mu}}{d\tau} = \frac{k^{*}_{\nu}}{\ms} F^{\mu\nu}
+\partial^\mu \ms \quad . \label{12} \eeqa Here $q^\mu = (t,{\bf
q})$ are the coordinates in Minkowski space and $F^{\mu\nu} =
\partial^\mu  V^\nu -
\partial^\nu V^\mu $ is the field strength tensor for $K^+$. For
$K^-$ where the vector field changes sign. The equation of motion
are identical, however, $F^{\mu\nu}$ has to be replaced by
$-F^{\mu\nu}$. The structure of Eqs. (12) may become more
transparent considering only the spatial components \beq \frac{d
{\bf k^*}}{d t} = - \frac{\ms}{E^*} \frac{\partial \ms }{\partial
{\bf q}} \mp \frac{\partial V^0 }{\partial {\bf q}} \pm \frac{{\bf
k}^*}{E^*} \times \left( \frac{\partial}{\partial {\bf q}} \times
{\bf V} \right) \label{13} \eeq where the upper (lower) signs refer
to $K^+$ ( $K^-$). The term proportional to the spatial component of
the vector potential gives rise to a momentum dependence which can
be attributed to a Lorentz force, i.e. the last term in Eq. (13).
Such a velocity dependent $({\bf v} = {\bf k}^* / E^* )$ Lorentz
force is a genuine feature of relativistic dynamics as soon as a
vector field is involved.

For the nuclear forces we use the standard momentum dependent Skyrme
interactions corresponding to a soft EOS (the compression modulus K
is 200 MeV). For the determination of the kaon mean field we adopt
the corresponding covariant scalar--vector description of the
non-linear $\sigma\omega$ model. Here we use the parametrization of
Ref.[23,24]  which corresponds to identical soft nuclear EOSs. The
shift of the production thresholds of the kaons by the in-medium
potentials is taken into account as described in[22,25] . The
hyperon fields are scaled according to SU(3) symmetry.

The flow of K$^+$ mesons presented in the Ref.[12] has already  been
simulated with Isospin Dependent Quantum Molecular Dynamics model
(IQMD) and Hadron String Dynamics(HSD) transport model and a good
agreement of the theoretical calculations with the data seem to be
obtained. Here we examine the consistency of the prediction of the
covariant kaon dynamics with the data. Before looking at our
calculated results with the model, it is worth to recall that there
are some distinctions between this model and IQMD though both of
them are based on the chiral mean field approximation for the kaon
in-medium interaction. Firstly, the way of kaons evolution in
nuclear medium is not the same, namely, in covariant kaon dynamics
the motion of the kaon meson is governed by covariant dynamical
equation mentioned above but in the IQMD model the kaon evolves
under the static potential given in Equ.(11)  and thus the
corresponding motion is not covariant. this is the most important
differences of these models. Secondly, the paramterization of kaon's
effective field is distinct, too. In IQMD the KN interaction
coefficient $\Sigma_{\mathrm{KN}} = 350 $MeV and the pion decay
constant $f_{\pi}=0.93$MeV. This scenario of the kaon's effective
field parametrization is called the Ko and Li scheme(KLP) [20] which
corresponds to the kaon potential $U_{K^{+}}(\rho _{0}) \approx
20MeV$ at saturation nuclear density. However, in the covariant kaon
dynamics, following Ref. [14], the Brown $\&$ Rho parametrization
(BRP): $ \Sigma_{\mathrm{KN}}$ = 450 MeV, $f_{\pi}^{*2}$ =
0.6$f_{\pi}^{2}$ for the vector field and $f_{\pi}^{*2}$ =
$f_{\pi}^{2}$ for the scalar part given by
$-\Sigma_{\mathrm{KN}}/f_{\pi}^{*2}\rhos $ are used. This scenario
accounts for the fact that the enhancement of the scalar part using
$f_{\pi}^{*2}$ is compensated by higher-order corrections in the
chiral expansion[2,4]. Up to saturation density the Brown and Rho
potential is $U_{K^{+}}(\rho _{0}) \approx 30MeV$. Additionally, the
other  pronounced difference between them is the treatment for
proton and neutron and the nuclear field, that is, in the former the
proton and neutron are taken as different kind of nucleons then the
isospin degree of freedom and the relevant symmetry energy  are
embed in the mean field of nucleons while in the later both of them
are considered as a same kind of particles and thereby the symmetry
energy is not considered. Keep these distinctions in mind we proceed
to inspect the calculation results here.

In Fig.1 we depict the variation of  K$^+$'s directed flow, $v_{1}$,
with respect to the rapidity of kaon produced in $^{58}_{28}$Ni +
$^{58}_{28}$Ni collisions at 1.91 AGeV. In the figure the line with
diamonds indicate the experiments data[12] and the curve with
solid-circles(hollow-circles) stand for the prediction of the
covariant kaon dynamical model with BRP(KLP). And the curve with
solid-triangles(hollow-triangle) represent the results of the QMD
model with static potentials of kaons, i.e., without the Lorentz
forces, which is incovariant for kaon motion equation  with
BRP(KLP).

\begin{center}
Fig.1
\end{center}

From this figure we can see that (1) the covariant kaon dynamics can
reproduce the data in which the Lorentz force(LF)  play a essential
role in determining the K$^+$'s directed  flow no matter if the KLP
or BRP parameterization scheme being taken into account in the
covariant kaon dynamics. This feature indicates, as has concluded in
Ref.[14] by evaluating the transverse flow $<p_{x}/m_{k}>$, that the
effect of the LF contribution in the covariant kaon dynamics pulls
the kaons back to the spectator matter; (2) the covariant kaon
dynamics with BRP parameterization give a more reasonable pattern of
the K$^+$'s directed  flow since there are discrepancies in the
cases of BRP and KLP when a same evolution equation being taken.
Comparatively, the results with BRP is slightly higher than that
with KLP. In view of the distinction induced by various
parameterization scenarios is subtle£¬we omit the results calculated
with KLP in following figures and concentrate our attention on the
influence of Lorentz-like force.

Fig.2 is  plot for the different directed flow of K$^+$'s in the
reaction systems at impact parameters $b=4.54$ fm(peripheral) and
$b=2.11$ fm(central). The dotted-curves with hollow-circles
corresponding to the variation of the directed flow with respect
to the transverse momentum $P_{t}$ of the kaons produced in the
reaction without the kaon final interaction being taken into
account and the other indicators of the curves are the same as
those appearing in Fig.1. The curves in the figure corresponding
respectively to the calculations via the covariant kaon
dynamics(with LF) and the evaluation without the kaon final
interaction(without $U_{K}$) are close to each other. This tell us
that the Lorentz force counterbalance the repulsive interaction of
the $K^+$ meson coming from nuclear environment and then resulting
in the predictions comparable  to the data. Therefore, the Lorentz
force, a genuine feature of the relativistic kaon dynamics as soon
as a vector field being involved, is also a crucial ingredients in
determining the K$^+$'s different directed flow.

\begin{center}
Fig.2
\end{center}

The rapidity dependence of  $V_{2}$  for K$^+$'s is plotted in the
Fig.3. The indicators of the curves are the same as figures 2. As
shown in the Fig.3 of Ref.[12] by comparing the predictions of IQMD
and HSD, the out-plane-flow $V_{2}$ is not sensitive to the kaon
final state interaction. This property is also demonstrated  in the
results of covariant kaon dynamics model.

\begin{center}

Fig.3

\end{center}

In summary, in the present work the directed and elliptic flows of
positively charged Kaon produced in  $^{58}_{28}$Ni +
$^{58}_{28}$Ni reaction at incident kinetic energy 1.91 AGeV
 are evaluated  within the covariant kaon dynamics combining by the quantum molecular dynamics
model. Our results show that it is necessary to include the Lorentz
force for reproducing the newly-delivered experiment data. The
sensitivity of $K^+$ the directed as well as differential directed
flow near target rapidity are observed by comparing the results
obtained via the covariant kaon dynamics model in the cases of with
or without kaon's final state potential including Lorentz force. And
such sensitivity can be attributed to the presence of Lorentz force
as soon as the space-like part of vector potential of kaon being
involved.

\hspace{2cm}

This work is supported in part by the National Natural Science
Foundation of China (NSFC) under Grant No. 11265013.


\newpage


\begin{figure}[tpb]


\vspace{0.1in} \caption{the variation of  K$^+$'s directed flow,
 with respect to the rapidity of kaon produced in
$^{58}_{28}$Ni + $^{58}_{28}$Ni collisions at incident energy 1.91
AGeV. theoretical results is obtained at impact parameter
$b=3.9$fm. }
 \label{Fig1}

\end{figure}

\begin{figure}[tpb]
\vspace{0.1in} \caption{ The different direct flow of K$^+$'s in
 $^{58}_{28}$Ni + $^{58}_{28}$Ni collisions at 1.91 AGeV with impact
parameters $b=4.95$fm(peripheral) and $b=2.11$fm(central).}
\label{Fig2}
\end{figure}

\begin{figure}[tpb]


\vspace{0.1in} \caption{Rapidity dependence of elliptic flow of
$K^+$'s in comparison to the predictions of covariant kaon
dynamics in the cases of with and without the Lorentz force.}

\label{Fig3}
\end{figure}

\end{document}